\documentclass[aps,prl,preprint,showpacs,amsmath,amssymb]{revtex4}

\usepackage{graphicx}
\usepackage{dcolumn}
\usepackage{bm}
\usepackage{textcomp,mathcomp}

\begin{document}

\title{Dynamics of reflection of ultracold atoms from a periodic 1D magnetic lattice potential}

\author{Mandip Singh, Russell McLean, Andrei Sidorov and Peter Hannaford}
\email{phannaford@swin.edu.au}
\affiliation{
\\Centre for Atom Optics and Ultrafast Spectroscopy and
\\ ARC Centre of Excellence for Quantum-Atom Optics
\\ Swinburne University of Technology, Melbourne, 3122, Australia}

\date{\today}

\begin{abstract}
We report on an experimental study of the dynamics of the reflection of ultracold atoms from a periodic one-dimensional magnetic lattice potential. The magnetic lattice potential of period 10~\textmu m is generated by applying a uniform bias magnetic field to a microfabricated periodic structure on a silicon wafer coated with a multilayered TbGdFeCo/Cr magneto-optical film. The effective thickness of the magnetic film is about 960~nm. A detailed study of the profile of the reflected atoms as a function of externally induced periodic corrugation in the potential is described. The effect of angle of incidence is investigated in detail. The experimental observations are supported by numerical simulations.
\end{abstract}


\maketitle


The control offered by optical lattices provides a rich platform to explore the physics of ultracold gases in periodic potentials. Optical lattices, which are based on the optical dipole force, have been used to realize coherent spin-dependent transport of atoms \cite{mandel:03} and multi-particle entanglement \cite{mandel2:03}. Many interesting phenomena which were earlier theoretically studied in condensed matter physics have now been experimentally explored using ultracold atoms in optical lattices \cite{grein02}.
 Another approach to realize periodic potentials utilizes a magnetic lattice which is based on the magnetic dipole force.  A magnetic lattice is a periodic array of magnetic traps which can be created either by structures of current carrying wires \cite{yinwire:02, grabowire:03} or permanent magnetic films fabricated on a substrate \cite{sinclair2:2005, ghanbari:2006, singh:2008}. Since magnetic lattices are based on miniaturized structures they are well suited to integration on a micro-chip. Permanent magnetic films offer relatively high magnetic field gradients and curvatures without any resistive heating on the chip. In addition, the thin structure and high electrical resistance of magnetic films can suppress thermal fluctuations, thereby offering higher trap stability. One-dimensional periodic magnetic structures produce a magnetic field that decays exponentially from the surface and which can be used to reflect atoms in weak field-seeking states \cite{opat92}. Such structures have been used to realize magnetic mirrors for atoms in weak field-seeking states \cite{roach95, sidorov96} and to manipulate atoms from a corrugated reflector \cite{rosen2000}. In recent experiments on a one-dimensional permanent magnetic lattice, radial trap frequencies of up to 90 kHz have been measured for $^{87}$Rb atoms trapped in the lattice at a distance close to 5~\textmu m from the surface \cite{singh:2008}. In another experiment a miniaturized structure of current carrying wires on a microchip  has been used to diffract a Bose Einstein condensate\cite{gunther05,gunther07}. In addition, interesting experiments on the manipulation of ultracold atoms in a two-dimensional array of magnetic microtraps have recently been reported \cite{gerritsma:2007}. In this paper we present a detailed study of the dynamics of the reflection of ultracold atoms from a permanent magnetic lattice potential with an externally controlled periodic corrugation.

\section{Magnetic field from a magnetic lattice structure}
\begin{center}
\begin{figure}
\begin{center}
\includegraphics[scale=0.6]{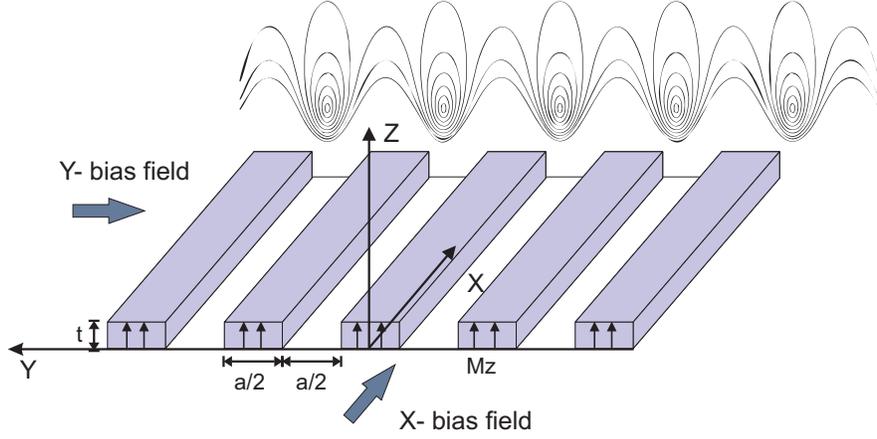}
\caption{\label{fig:lattice}Schematic showing an array of perpendicularly magnetized parallel slabs with period $a$. A one-dimensional magnetic lattice of elongated microtraps with non-zero potential minima is formed by applying bias fields along the $x$- and $y$-directions.}
\end{center}
\end{figure}
\end{center}

 In order to produce the required potential a grooved silicon microstructure of period 10~\textmu m was coated with multilayered Tb$_{10}$Gd$_{6}$Fe$_{80}$Co$_{4}$/Cr magneto-optical film \cite{wang2:05}. Such a structure when perpendicularly magnetized generates periodically varying magnetic field components and by applying a uniform bias magnetic field (along the $y$-direction) to the structure an array of magnetic traps can be produced \cite{ghanbari:2006}. A schematic of a periodic magnetic structure of period $a$, thickness $t$ and perpendicular magnetization $M_{z}$ is shown in Fig.~\ref{fig:lattice}.

For an infinite array structure the components of the magnetic field along the $y$- and $z$-directions are given by \cite{ghanbari:2006}

 \begin{equation}
      \label{eq:2.B1}
      B_{y}=B_{s}[(1-e^{-k t})e^{-k[z-t]}\sin(ky)-\frac{1}{3}(1-e^{-3k t})e^{-3 k[z-t]}\sin(3 k y)+\cdot\cdot\cdot]+ B_{by}
      \end{equation}
               \begin{equation}
      \label{eq:2.B2}
      B_{z}=B_{s}[(1-e^{-k t})e^{-k[z-t]}\cos(ky)-\frac{1}{3}(1-e^{-3k t})e^{-3 k[z-t]}\cos(3 k y)+\cdot\cdot\cdot]+ B_{bz}
      \end{equation}

         where $k = 2\pi/{a}$,  $B_{s}= 4M_{z}$ (Gaussian units), $M_{z}$ is the magnetization (assumed to be in the perpendicular direction) and $B_{bi}$ for $i= x,y,z$ are the bias field components along $x$, $y$, $z$-directions, respectively.

       For distances from the surface large compared to $ a/4\pi$ the effect of higher spatial harmonics is negligible and the field components can be written as

      \begin{equation}
      \label{eq:Bc1}
      B_{x}= B_{bx}
      \end{equation}
       \begin{equation}
       \label{eq:Bc2}
      B_{y} = B_{0}\sin(ky)e^{-kz} + B_{by}
      \end{equation}
       \begin{equation}
      \label{eq:Bc3}
      B_{z} = B_{0}\cos(ky)e^{-kz} + B_{bz}
        \end{equation}
 where  $ B_{0}= B_{s}(e^{kt}-1)$.
  In the absence of bias fields the magnitude of the magnetic field is given by

      \begin{equation}
      \label{eq:Bc4}
      B(x,y)= B_{0}e^{-kz}
      \end{equation}
i.e., the magnitude of the field decreases exponentially with distance $z$ from the surface. Thus the field gradient repels atoms in weak field-seeking states and such a structure behaves as a magnetic mirror \cite{opat92}. In the absence of bias magnetic fields $B_{by}$ and $B_{bz}$ the intensity of the magnetic field generated by the magnetic structure is uniform in the $x$-$y$ plane. However, by applying a bias field in the $y$- or $z$-direction one can induce a periodic corrugation in the field magnitude along the $y$-direction. Such an exponentially increasing corrugated potential can significantly affect the spatial profile of an ultracold cloud of atoms projected towards it. The main objective of this paper is to study the detailed dynamics of an ultracold cloud reflected from the periodic corrugated potential.

\section{Atoms in the corrugated magnetic potential}
In order to study the profile of the ultracold cloud reflected or released from the potential we start by evaluating the forces acting on an atom in the potential. In the case where the spatial size of an atomic wavepacket moving in a magnetic field $B(x,y,z)$ is much smaller than the corrugation period, the atom can be treated as a classical point object. The components of the force acting on the atom in the magnetic lattice potential with gravity ($g$) along the $z$-direction can then be expressed as
            \begin{equation}
      \label{eq:b5}
      m\frac{\mathrm d^{2}x}{\mathrm d t^{2}} = - m_{F}g_{F}\mu_{B} \left(\frac{\partial B(x,y,z)}{\mathrm \partial x}\right)
           \end{equation}
                \begin{equation}
      \label{eq:b6}
     m\frac{\mathrm d^{2} y}{\mathrm d t^{2}} = - m_{F}g_{F}\mu_{B} \left(\frac{\partial B(x,y,z) }{\mathrm \partial y}\right)
           \end{equation}
                \begin{equation}
      \label{eq:b7}
      m\frac{\mathrm d^{2}z}{\mathrm d t^{2}} = - m_{F}g_{F}\mu_{B}\left( \frac{\partial B(x,y,z) }{\mathrm \partial z}\right)+ m g
           \end{equation}
\begin{center}
\begin{figure}
\begin{center}
\includegraphics[scale=.55]{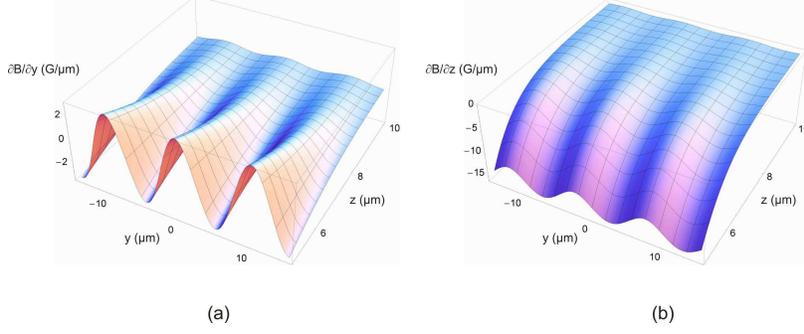}
\caption{\label{fig:derivative} Plots showing the partial derivatives of the magnetic field magnitude in the lattice potential (a) with respect to $y$ and (b) with respect to $z$. The parameters used in the calculations are given in the text.}
\end{center}
\end{figure}
\end{center}
         where $m$ is the mass of the atom,  $m_{F}$ is the magnetic quantum number, $g_{F}$ is the Land\'{e} g-factor and $\mu_{B}$ is the Bohr magneton. From Eqs.~\ref{eq:Bc2} and~\ref{eq:Bc3} the magnitude of the magnetic field in the presence of uniform bias fields is given by

          \begin{equation}
         \label{eq:b8}
          B(x,y,z) = [B_{bx}^{2}+B_{by}^{2} + B_{bz}^{2}+ B_{0}^{2} e^{-2kz} + 2 B_{0} e^{-kz}(B_{bz}\cos(ky)+B_{by}\sin(ky))]^{1/2}
           \end{equation}
           and the partial derivatives are

             \begin{equation}
         \label{eq:b9}
         \frac{\partial B(x,y,z)}{\partial x}= 0
           \end{equation}

               \begin{equation}
         \label{eq:b10}
         \frac{\partial B(x,y,z)}{\partial y}= \frac{k B_{0}e^{-kz}(B_{by}\cos{ky}-B_{bz}\sin{ky})}{B(x,y,z)}
           \end{equation}

                     \begin{equation}
         \label{eq:b11}
         \frac{\partial B(x,y,z)}{\partial z}= - \frac{k B_{0}e^{-2kz}(B_{0} + e^{kz}(B_{bz}\cos{ky}+ B_{by}\sin{ky}))}{B(x,y,z)}
           \end{equation}
          Inserting  Eqs.~\ref{eq:b9},~\ref{eq:b10} and~\ref{eq:b11} into Eqs.~\ref{eq:b5},~\ref{eq:b6} and~\ref{eq:b7} and solving, the trajectory of an atom in the magnetic potential can be calculated. Profiles of $\partial B(x,y,z)/\partial y $ and $\partial B(x,y,z)/\partial z$ in the $x = c$ plane (where $c$ is a constant) are shown in Fig.~\ref{fig:derivative} (a) and (b), respectively, for $B_{bx}$~=~45~G, $B_{by}$~=~8~G, $B_{bz}$~=~0~G, $t= 1$~\textmu m, $a = 10$~\textmu m and $4\pi M_{z}=$~3~kG. Figure.~\ref{fig:derivative} (a) indicates that the $y$-component of the force, which is proportional to $\partial B(x,y,z)/\partial y $, is an oscillatory function of $y$ and provides horizontal momentum to the atoms while the negative values of $\partial B(x,y,z)/\partial z$ cause repulsion from the lattice (for atoms in weak field-seeking states). For the one-dimensional potential considered, $\partial B(x,y,z)/\partial x$~=~0, which results in a zero force along the $x$-axis.

\section{Experiment: Effect of corrugation}
\label{corrg}
\begin{center}
\begin{figure}
\begin{center}
\includegraphics[scale=0.6]{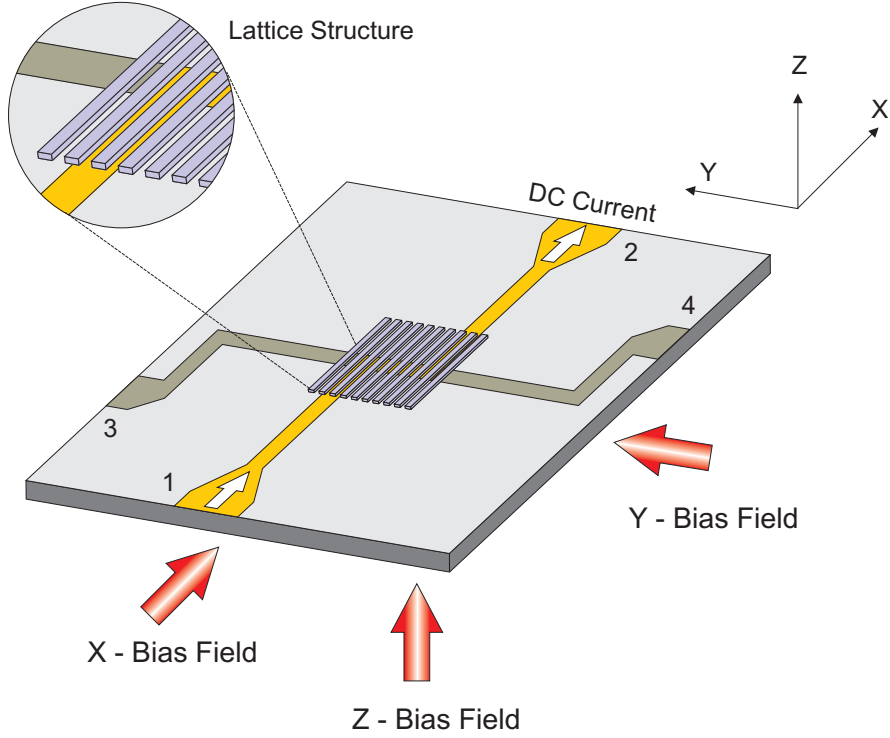}
\caption{\label{fig:wirechip} Periodic magnetic structure and current carrying wires.}
\end{center}
\end{figure}
\end{center}

       The experimental setup is based on a hybrid atom chip \cite{whitlock:06, singh:2008} which consists of a permanent magnetic structure and current-carrying wires. The permanent magnetic structure was produced by coating a multilayered magneto-optical film (Tb$_{10}$Gd$_{6}$Fe$_{80}$Co$_{4}$/Cr) of effective thickness 960~nm on a grooved Si structure. The period of the (10~mm $\times$10~mm) grooved structure is 10~$\mu$m. The coated structure was perpendicularly magnetized with $4 \pi M_{z} \approx$~3~kG. Measurements indicate the coercivity of the film was about 6~kOe and the Curie temperature is normally about 300$^{0}$C. The grooved Si wafer was then positioned on the current-carrying wire structure. The position of the permanent magnetic structure with respect to the wires is shown schematically in Fig.~\ref{fig:wirechip}, where gravity is acting along the $z$-direction (the chip is mounted face down).

          The periodic corrugation in the magnetic potential was introduced externally by applying a bias field $B_{by}$ and keeping $B_{bz}$~=~0. The experimental protocol is as follows. About $2 \times 10^{5}$ $^{87}$Rb atoms in the weak field-seeking state $|F=2, m_{F}=2\rangle$ were cooled close to the BEC transition by evaporative cooling in a Z-shaped wire (pin numbers 1, 2 in Fig.~\ref{fig:wirechip}) magnetic trap located 150-250 $\mu$m from the surface of the structure. At this distance the exponentially decaying magnetic field from the 10~\textmu m period permanent magnetic structure is negligibly small. After the evaporative cooling stage the externally applied magnetic field along the $y$-axis (which allows adjustment of the trap offset in the Z-wire trap) was linearly reduced to zero in 10 ms keeping $I_{z}$~=~35~A (pin numbers 1, 2 in Fig.~\ref{fig:wirechip}) and $B_{bx}$~=~45~G. After waiting another 10~ms the value of $I_{z}$ was linearly reduced from 35~A to 22~A in 1.4~ms. Immediately after this 1.4~ms interval $I_{z}$ was quickly reduced to zero and the corrugation inducing bias field $B_{by}$ was simultaneously switched to the desired value. Both $I_{z}$ and $B_{by}$ were able to be switched on (off) in less than 1~ms. The process of decreasing $I_{z}$ in 1.4~ms moves the trap centre towards the periodic magnetic lattice potential and therefore the trapped ultracold cloud is propelled upwards towards the lattice potential with a velocity $v_{z}$ in the range -100~mm/s to -130~mm/s. When $I_{z}$ was switched off the Z-trap disappeared but the ultracold cloud continued to move in the presence of $B_{bx}$~=~45~G and $B_{by}$. The moving cloud enters the exponentially increasing magnetic field of the magnetic lattice and experiences an exponentially increasing repulsive force in the $z$-direction. In the case of an external applied bias field $B_{by}$ the $y$-component of the force on the atoms due to the nonzero value of $\partial B(x,y,z)/\partial y$ also spreads them horizontally (in the $y$-direction).

\begin{center}
\begin{figure}
\begin{center}
\includegraphics[scale=0.5]{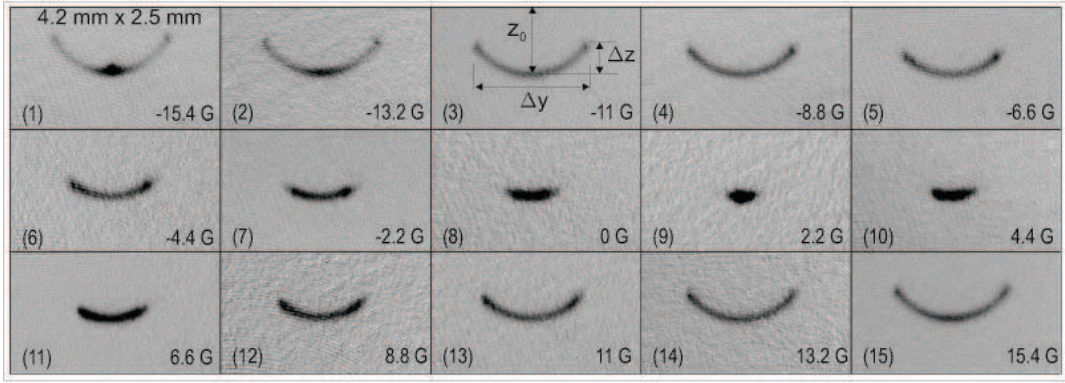}
\caption{\label{fig:bycorexp} Absorption images showing experimentally observed spatial profiles of atoms reflected from a corrugated potential for different values of applied bias field $B_{by}$ (as indicated in the bottom right corner of each image) and hence different amplitudes of corrugation (for 9~ms time of flight). The horizontal and vertical directions correspond to the $y$- and $z$-axis, respectively. Gravity is in the downward direction, and the chip surface is above the top of each image.}
\end{center}
\end{figure}
\end{center}
\begin{center}
\begin{figure}
\begin{center}
\includegraphics[scale=1.2]{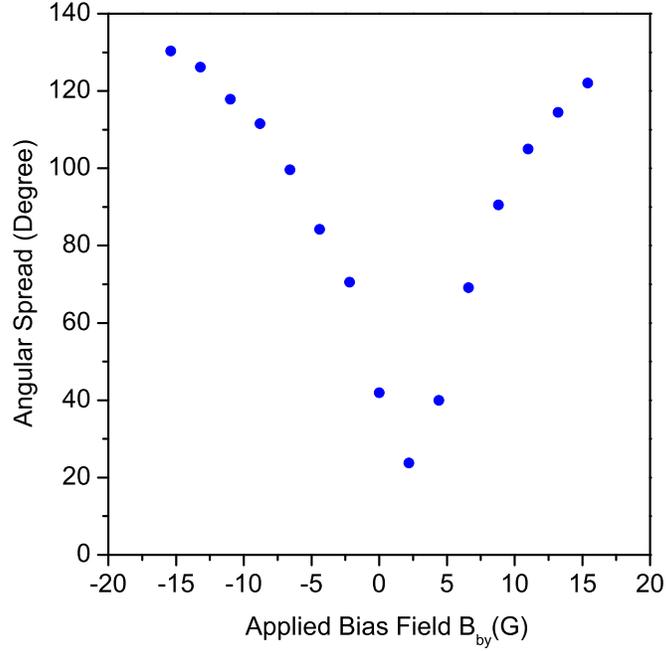}
\caption{\label{fig:byplot} Angular spread of atoms as a function of applied bias field $B_{by}$. The minimum of the angular spread occurs at $B_{by}\simeq$~2.2~G rather than at $B_{by}$~=~0~G due the presence of a stray field.}
\end{center}
\end{figure}
\end{center}

            Absorption images were taken after the atom cloud had interacted with the periodic corrugated potential.  The amplitude of the corrugation is varied by changing the value of $B_{by}$. A set of images is shown in Fig.~\ref{fig:bycorexp} for a 9 ms time of flight for a range of values of $B_{by}$ where the atomic profiles closely resemble a circular arc. It is evident from Fig.~\ref{fig:bycorexp} that as the magnitude of $B_{by}$ and hence the corrugation is reduced the angular spread of the reflected atoms decreases. The angular spread is given by the angle subtended by the circular arc at the centre of the circle and is equal to $2 \arctan (\Delta y/2(z_{0}-\Delta z))$ for the horizontal ($\Delta y$) and the vertical ($\Delta z$) width of the profile shown in Fig.~\ref{fig:bycorexp}. The radius of the circular arc $z_{0}\approx$~1.4~mm . A plot showing the angular spread of the released cloud as a function of $B_{by}$ is presented in~Fig.~\ref{fig:byplot}. It is evident from Fig.~\ref{fig:byplot} that the minimum angular spread appears at $B_{by}$~=~2.2~G, instead of $B_{by}$~=~0. This offset is due to the presence of a stray field of the same order which cancelled the applied field $B_{by}$. This experiment establishes that the angular spread  originates from the periodic corrugation in the magnetic field pattern that can be controlled externally. In the case of an evanescent light wave mirror it has been shown that the effect of roughness can also produce curvature in the profile of the reflected atomic cloud \cite{perrin06}.

\section{Simulation: Effect of corrugation}
\begin{center}
\begin{figure}
\begin{center}
\includegraphics[scale=0.42]{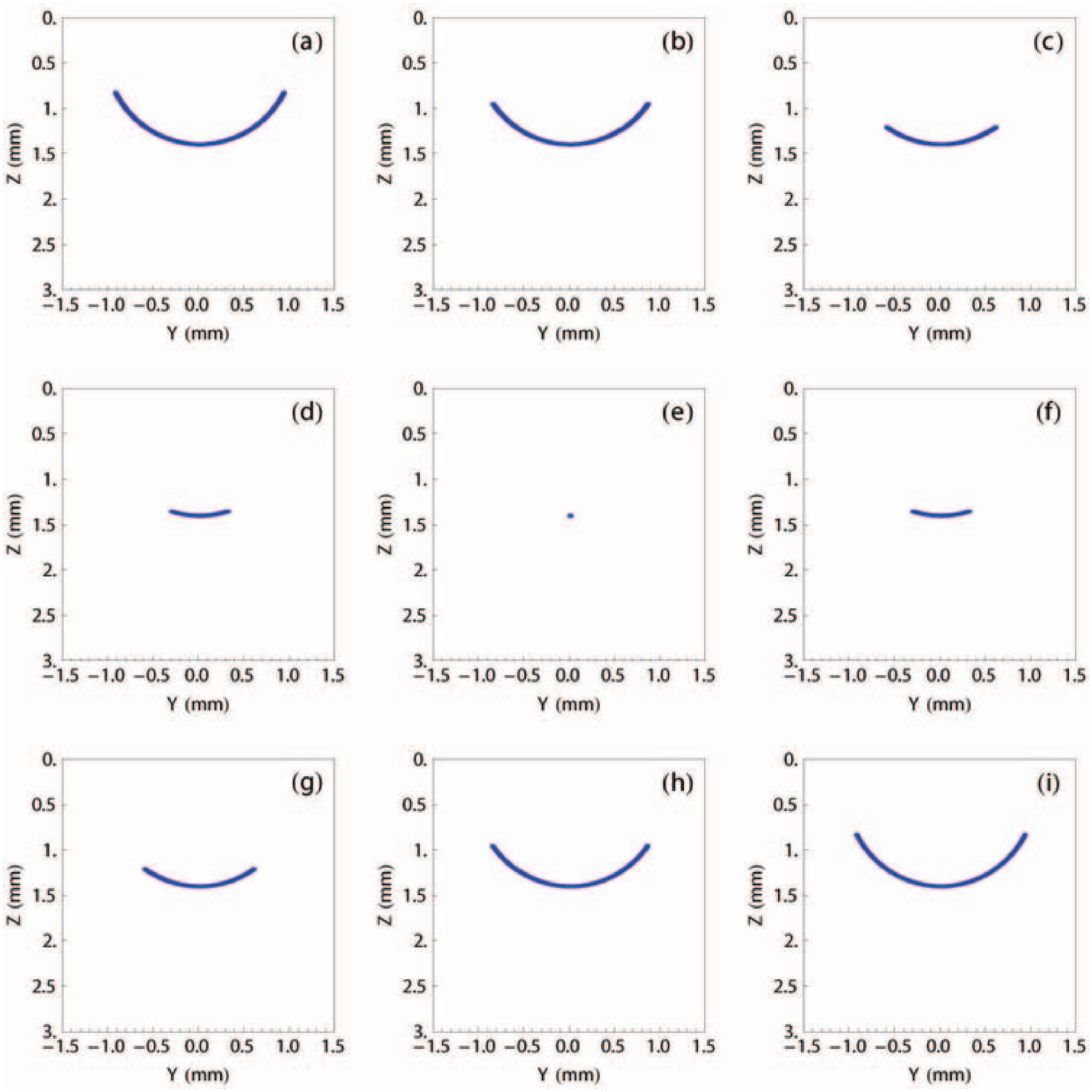}
\caption{\label{fig:bytheory} Calculated spatial profiles of the atoms as a function of corrugation. $B_{by}=$ (a)~-~6.6~G. (b)~-~4.4~G. (c)~-~2.2~G. (d)~-~1~G. (e)~0~G. (f)~1~G. (g)~2.2~G. (h)~4.4~G. (i)~6.6~G.}
\end{center}
\end{figure}
\end{center}

In order to verify the interpretation of the above experimental observations the differential equations ~\ref{eq:b5},~\ref{eq:b6} and~\ref{eq:b7} were solved numerically for different values of $B_{by}$. In the calculations the magnetic structure parameters  $a$~=~10 $\mu$m, $4 \pi M_{z}$~=~3~kG and $t$~=~1~$\mu$m were used. The calculations were performed by taking the initial velocity (at t=0) $v_{z}$~=~-125~mm/s, the initial position of the cloud (at t=0) $z_{i}=$~60~$\mu$m, $B_{bx}$~=45~G, $B_{bz}$~=~0~G and zero temperature. An initial cloud width of 30~$\mu$m along the $y$-direction was assumed which is three times the lattice period. A set of calculated profiles of the cloud as a function of $B_{by}$ is shown in~Fig.~\ref{fig:bytheory} which indicates that the angular spread gradually reduces as the corrugation is reduced to zero, as observed in the experiment.

\section{Varying the Angle of incidence}
 In the previous experiment the ultracold cloud was projected perpendicular to the plane of the magnetic structure, \emph{i.e.}, $v_{x}$ = 0, $v_{y}$ = 0, and only $v_{z}$ was nonzero. For an ultracold cloud moving towards the lattice in the $y$-$z$ plane the angle of incidence with respect to the $z$-axis is $\theta = \arctan (v_{y}/v_{z})$ which was zero in the previous experiment. In this experiment the effect of angle of incidence is studied. The angle of incidence was varied by varying $v_{y}$ and keeping $v_{z}$ constant.
\begin{center}
\begin{figure}
\begin{center}
\includegraphics[scale=0.5]{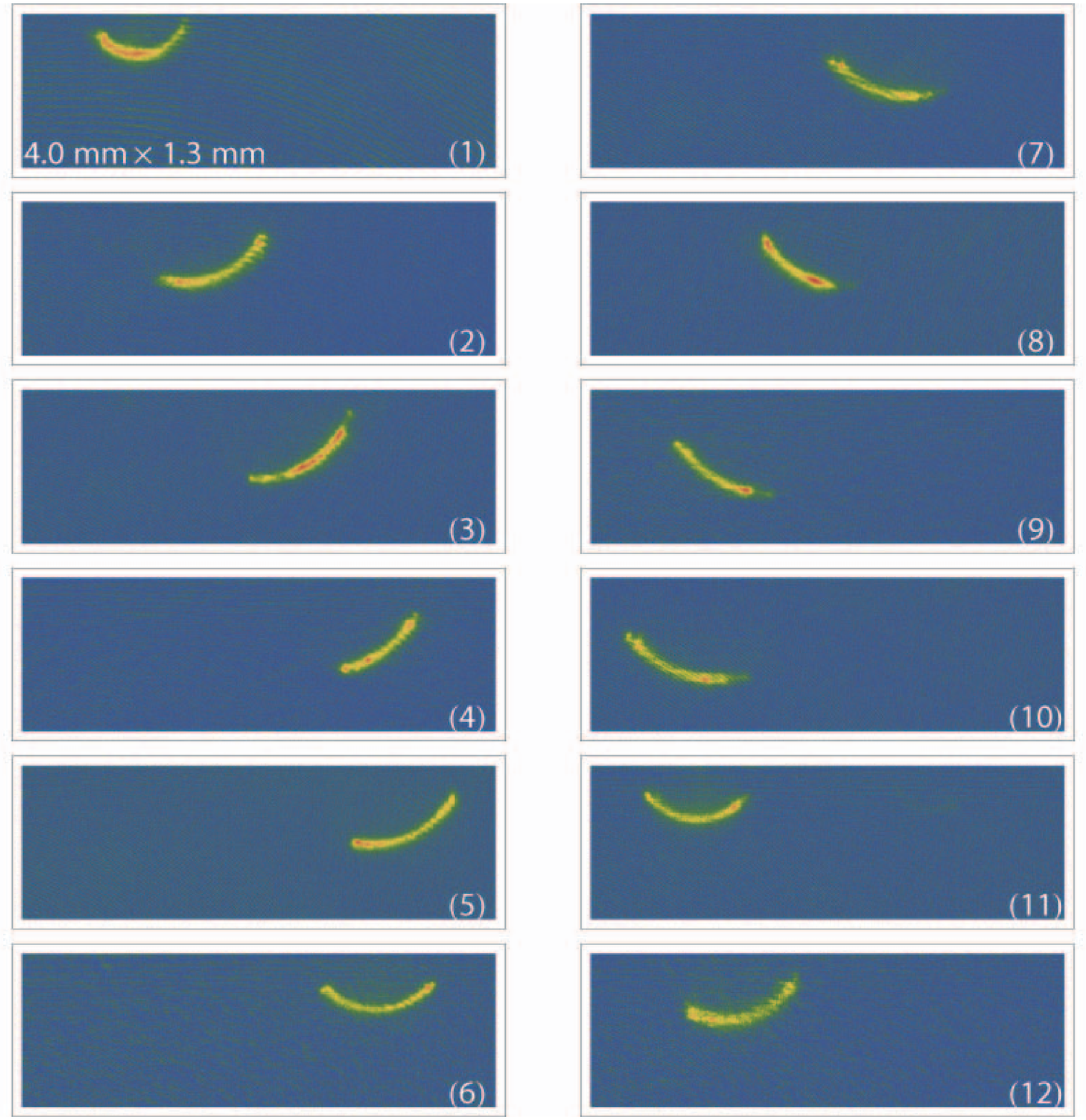}
\caption{\label{fig:grazexp} Series of absorption images showing the atomic density profile of reflected atoms for different values of $t_{hold}$. The time step in successive measurements is 4 ms. The horizontal and vertical directions correspond to the $y$- and $z$-axis, respectively.}
\end{center}
\end{figure}
\end{center}

               In order to provide a velocity component $v_{y}$, the ultracloud cloud was made to oscillate in the $y$-direction in the Z-trap before moving towards the lattice. The protocol used to project the cloud at an angle with respect to the $z$-axis is as follows. After completion of the evaporative cooling stage an ultracold cloud of $2.5 \times 10^{5}$ atoms close to the critical temperature was obtained in the Z-wire trap. Immediately after evaporation  $B_{by}$ was decreased linearly to zero in 10 ms. In the next 50 ms $I_{z}$~=~35~A was decreased to 10~A and the current ($I_{zs}$) in the semi-long Z-wire (pin numbers 1, 4 in Fig.~\ref{fig:wirechip}) was increased from 0~A to 25~A synchronously such that $I_{z} + I_{zs}$~=~35~A. In the next 10 ms both $I_{z}$ and $I_{zs}$ were restored linearly to their initial values of 35~A and 0~A, respectively, where $B_{bx}$~=~45~G was constant throughout. This process provides a kick to the cloud along the $y$-direction in the Z-trap and the cloud starts to oscillate. Therefore, $v_{y}$ also oscillates back and forth. This is the first part of the protocol. In the second part the cloud was moved towards the lattice after holding for a variable time $t_{hold}$ while it was oscillating. The cloud was moved towards the lattice by linearly decreasing $I_{z}$ from 35~A to 17~A in 15~ms. After holding $I_{z}$ at 17~A for 0.3 ms it was quickly reduced to zero. The atoms reflected from the corrugated potential were detected through absorption imaging after 6~ms time of flight. The corrugation in this experiment originates from the magnetic field generated by the current through the Z-wire. Measurements were taken for different values of $t_{hold}$ to vary $v_{y}$ of the cloud (at the time when it interacts with the magnetic potential).

\begin{center}
\begin{figure}
\begin{center}
\includegraphics[scale=0.5]{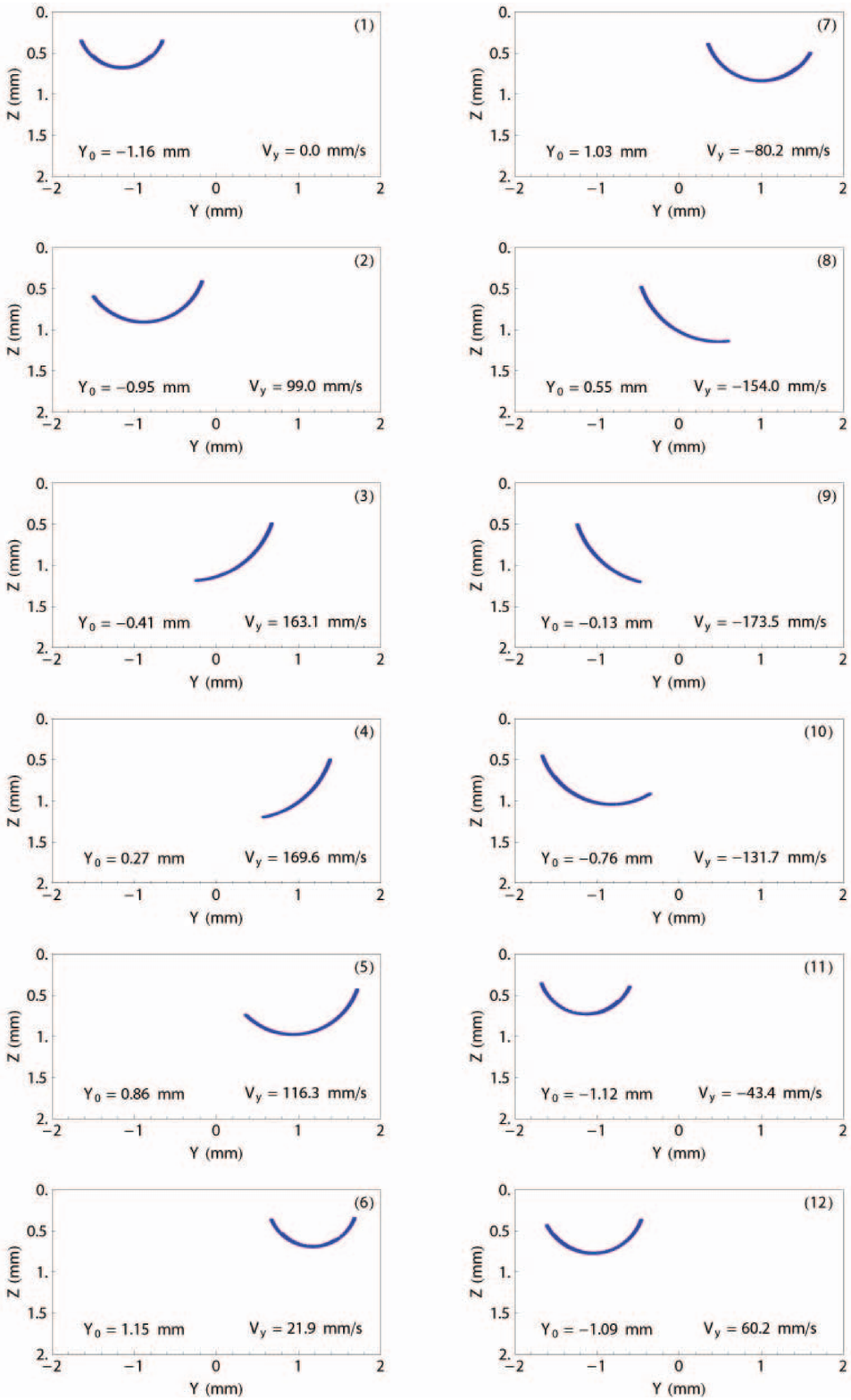}
\caption{\label{fig:grazth} Calculated spatial profiles of the reflected atoms as a function of axial position and axial velocity.}
\end{center}
\end{figure}
\end{center}
                  The axial position ($y_{0}$) and axial velocity ($v_{y}$) of the oscillating cloud at the time when it interacts with the lattice were evaluated by fitting a harmonic function to the axial oscillations. The axial position and axial velocity are given by $y_{0}=A_{0}\cos(\omega_{a} t_{hold}+\varphi)$ and $v_{y}=-\omega_{a}A_{0}\sin(\omega_{a}t_{hold}+\varphi)$, respectively. Thus, in this experiment the ultracold cloud can be made to interact with the lattice potential with different $v_{y}$ by varying $t_{hold}$. This also results in the interaction occurring at different locations.

                  In~Fig.~\ref{fig:grazexp} a series of absorption images of the reflected atoms is shown for different values of $t_{hold}$ over a full cycle of the axial oscillation where the time step in successive images is 4 ms. It is apparent from this data that the profile of the reflected atoms is tilted at an angle which depends on $v_{y}$ but is independent of the axial position $y_{0}$.  At the turning point of the axial oscillation the axial velocity $v_{y}$ is zero and therefore the angle of incidence ($\theta$) is also zero (normal incidence). The horizontal distance between the two turning points (located at the centre of each profile) when $\theta=0$ represents 2$A_{0}$. These two turning points correspond approximately to image numbers (1) and (6) in~Fig.~\ref{fig:grazexp} and the distance between them is 2$A_{0}$~=~2.32~mm. These profiles show that uniformity of the corrugation persists over many lattice periods.

                  The experimental results were verified by calculating the profile of reflected atoms for different values of $y_{0}$ and $v_{y}$ by solving Eqs.~\ref{eq:b5},~\ref{eq:b6} and~\ref{eq:b7}. The values of  $y_{0}$ and $v_{y}$ for each simulation were calculated from $y_{0}=A_{0}\cos(\omega_{a} t_{hold}+\varphi)$ and $v_{y}=-\omega_{a}A_{0}\sin(\omega_{a}t_{hold}+\varphi)$ for $\varphi=\pi$, $\omega_{a}=2\pi\times24$ rad/s in successive time steps of 4 ms. A series of calculated atomic profiles for the experimental parameters is shown in~Fig.~\ref{fig:grazth}, where the time of flight is 6 ms  and $v_{z }$~=~-102~mm/s for each plot.

\section{Summary and Conclusions}
               A detailed study and analysis of the reflection of an ultracold cloud of atoms from a one-dimensional magnetic lattice potential of period 10~\textmu m has been presented. The magnetic lattice was constructed on a grooved Si structure coated with perpendicularly magnetized multilayered TbGdFeCo/Cr magneto optical film and mounted on a hybrid atom chip. It has been shown experimentally and theoretically that the field induced corrugation can significantly change the profile of the cloud to a curved shape. In addition a detailed study of the effect of angle of incidence has been presented; in this case the atomic cloud interacts with different regions of the lattice where the extreme points are separated by up to about 2.3~mm. The observed atomic profiles have been explained using classical equations of motion applied to atoms in the magnetic lattice potential. The results provide new insights into the manipulation of ultracold atoms diffusively reflecting from a periodic corrugated potential. In future we would like to explore potential applications such as how the reflection from the periodic corrugated potential can be used to characterize the magnetic lattice potential and how the effect of periodic corrugation differs from a random corrugation.

\begin{acknowledgments}
We would like to thank Brenton Hall for useful discussions. This work is supported by the Australian Research Council Centre of Excellence for Quantum-Atom Optics and a Swinburne University Strategic Initiative Grant.
\end{acknowledgments}

\bibliography{mandip_diff}

\begin{thebibliography}{18}
\expandafter\ifx\csname natexlab\endcsname\relax\def\natexlab#1{#1}\fi
\expandafter\ifx\csname bibnamefont\endcsname\relax
  \def\bibnamefont#1{#1}\fi
\expandafter\ifx\csname bibfnamefont\endcsname\relax
  \def\bibfnamefont#1{#1}\fi
\expandafter\ifx\csname citenamefont\endcsname\relax
  \def\citenamefont#1{#1}\fi
\expandafter\ifx\csname url\endcsname\relax
  \def\url#1{\texttt{#1}}\fi
\expandafter\ifx\csname urlprefix\endcsname\relax\def\urlprefix{URL }\fi
\providecommand{\bibinfo}[2]{#2}
\providecommand{\eprint}[2][]{\url{#2}}

\bibitem[{\citenamefont{Mandel et~al.}(2003{\natexlab{a}})\citenamefont{Mandel,
  Greiner, Widera, Rom, H{\"{a}}nsch, and Bloch}}]{mandel:03}
\bibinfo{author}{\bibfnamefont{O.}~\bibnamefont{Mandel}},
  \bibinfo{author}{\bibfnamefont{M.}~\bibnamefont{Greiner}},
  \bibinfo{author}{\bibfnamefont{A.}~\bibnamefont{Widera}},
  \bibinfo{author}{\bibfnamefont{T.}~\bibnamefont{Rom}},
  \bibinfo{author}{\bibfnamefont{T.~W.} \bibnamefont{H{\"{a}}nsch}},
  \bibnamefont{and} \bibinfo{author}{\bibfnamefont{I.}~\bibnamefont{Bloch}},
  \bibinfo{journal}{Phys. Rev. Lett.} \textbf{\bibinfo{volume}{91}},
  \bibinfo{pages}{010407} (\bibinfo{year}{2003}{\natexlab{a}}).

\bibitem[{\citenamefont{Mandel et~al.}(2003{\natexlab{b}})\citenamefont{Mandel,
  Greiner, Widera, Rom, H{\"{a}}nsch, and Bloch}}]{mandel2:03}
\bibinfo{author}{\bibfnamefont{O.}~\bibnamefont{Mandel}},
  \bibinfo{author}{\bibfnamefont{M.}~\bibnamefont{Greiner}},
  \bibinfo{author}{\bibfnamefont{A.}~\bibnamefont{Widera}},
  \bibinfo{author}{\bibfnamefont{T.}~\bibnamefont{Rom}},
  \bibinfo{author}{\bibfnamefont{T.~W.} \bibnamefont{H{\"{a}}nsch}},
  \bibnamefont{and} \bibinfo{author}{\bibfnamefont{I.}~\bibnamefont{Bloch}},
  \bibinfo{journal}{Nature} \textbf{\bibinfo{volume}{425}},
  \bibinfo{pages}{937} (\bibinfo{year}{2003}{\natexlab{b}}).

\bibitem[{\citenamefont{Greiner et~al.}(2002)\citenamefont{Greiner, Mandel,
  Esslinger, H{\"{a}}nsch, and Bloch}}]{grein02}
\bibinfo{author}{\bibfnamefont{M.}~\bibnamefont{Greiner}},
  \bibinfo{author}{\bibfnamefont{O.}~\bibnamefont{Mandel}},
  \bibinfo{author}{\bibfnamefont{T.}~\bibnamefont{Esslinger}},
  \bibinfo{author}{\bibfnamefont{T.~W.} \bibnamefont{H{\"{a}}nsch}},
  \bibnamefont{and} \bibinfo{author}{\bibfnamefont{I.}~\bibnamefont{Bloch}},
  \bibinfo{journal}{Nature} \textbf{\bibinfo{volume}{415}}, \bibinfo{pages}{39}
  (\bibinfo{year}{2002}).

\bibitem[{\citenamefont{Yin et~al.}(2001)\citenamefont{Yin, Gao, Hu, and
  Wang}}]{yinwire:02}
\bibinfo{author}{\bibfnamefont{J.}~\bibnamefont{Yin}},
  \bibinfo{author}{\bibfnamefont{W.}~\bibnamefont{Gao}},
  \bibinfo{author}{\bibfnamefont{J.}~\bibnamefont{Hu}}, \bibnamefont{and}
  \bibinfo{author}{\bibfnamefont{Y.}~\bibnamefont{Wang}},
  \bibinfo{journal}{Opt. Commun.} \textbf{\bibinfo{volume}{206}},
  \bibinfo{pages}{99} (\bibinfo{year}{2001}).

\bibitem[{\citenamefont{Grabowski and Pfau}(2003)}]{grabowire:03}
\bibinfo{author}{\bibfnamefont{A.}~\bibnamefont{Grabowski}} \bibnamefont{and}
  \bibinfo{author}{\bibfnamefont{T.}~\bibnamefont{Pfau}},
  \bibinfo{journal}{Eur. Phys. J. D.} \textbf{\bibinfo{volume}{22}},
  \bibinfo{pages}{347} (\bibinfo{year}{2003}).

\bibitem[{\citenamefont{Sinclair et~al.}(2005)\citenamefont{Sinclair, Retter,
  Curtis, Hall, Garcia, Eriksson, Sauer, and Hinds}}]{sinclair2:2005}
\bibinfo{author}{\bibfnamefont{C.~D.~J.} \bibnamefont{Sinclair}},
  \bibinfo{author}{\bibfnamefont{J.~A.} \bibnamefont{Retter}},
  \bibinfo{author}{\bibfnamefont{E.~A.} \bibnamefont{Curtis}},
  \bibinfo{author}{\bibfnamefont{B.~V.} \bibnamefont{Hall}},
  \bibinfo{author}{\bibfnamefont{I.~L.} \bibnamefont{Garcia}},
  \bibinfo{author}{\bibfnamefont{S.}~\bibnamefont{Eriksson}},
  \bibinfo{author}{\bibfnamefont{B.~E.} \bibnamefont{Sauer}}, \bibnamefont{and}
  \bibinfo{author}{\bibfnamefont{E.~A.} \bibnamefont{Hinds}},
  \bibinfo{journal}{Eur. Phys. J. D.} \textbf{\bibinfo{volume}{35}},
  \bibinfo{pages}{105} (\bibinfo{year}{2005}).

\bibitem[{\citenamefont{Ghanbari et~al.}(2006)\citenamefont{Ghanbari, Kieu,
  Sidorov, and Hannaford}}]{ghanbari:2006}
\bibinfo{author}{\bibfnamefont{S.}~\bibnamefont{Ghanbari}},
  \bibinfo{author}{\bibfnamefont{T.~D.} \bibnamefont{Kieu}},
  \bibinfo{author}{\bibfnamefont{A.}~\bibnamefont{Sidorov}}, \bibnamefont{and}
  \bibinfo{author}{\bibfnamefont{P.}~\bibnamefont{Hannaford}},
  \bibinfo{journal}{J. Phys. B: At. Mol. Opt. Phys.}
  \textbf{\bibinfo{volume}{39}}, \bibinfo{pages}{847} (\bibinfo{year}{2006}).

\bibitem[{\citenamefont{Singh et~al.}(2008)\citenamefont{Singh, Volk, Akulshin,
  McLean, Sidorov, and Hannaford}}]{singh:2008}
\bibinfo{author}{\bibfnamefont{M.}~\bibnamefont{Singh}},
  \bibinfo{author}{\bibfnamefont{M.}~\bibnamefont{Volk}},
  \bibinfo{author}{\bibfnamefont{A.}~\bibnamefont{Akulshin}},
  \bibinfo{author}{\bibfnamefont{R.}~\bibnamefont{McLean}},
  \bibinfo{author}{\bibfnamefont{A.}~\bibnamefont{Sidorov}}, \bibnamefont{and}
  \bibinfo{author}{\bibfnamefont{P.}~\bibnamefont{Hannaford}},
  \bibinfo{journal}{J. Phys. B: At. Mol. Opt. Phys.}
  \textbf{\bibinfo{volume}{41}}, \bibinfo{pages}{065301}
  (\bibinfo{year}{2008}).

\bibitem[{\citenamefont{Opat et~al.}(1992)\citenamefont{Opat, Wark, and
  Cimmino}}]{opat92}
\bibinfo{author}{\bibfnamefont{G.~I.} \bibnamefont{Opat}},
  \bibinfo{author}{\bibfnamefont{S.~J.} \bibnamefont{Wark}}, \bibnamefont{and}
  \bibinfo{author}{\bibfnamefont{A.}~\bibnamefont{Cimmino}},
  \bibinfo{journal}{Appl. Phys. B.} \textbf{\bibinfo{volume}{54}},
  \bibinfo{pages}{396} (\bibinfo{year}{1992}).

\bibitem[{\citenamefont{Roach et~al.}(1995)\citenamefont{Roach, Abele, Boshier,
  Grossman, Zetie, and Hinds}}]{roach95}
\bibinfo{author}{\bibfnamefont{T.~M.} \bibnamefont{Roach}},
  \bibinfo{author}{\bibfnamefont{H.}~\bibnamefont{Abele}},
  \bibinfo{author}{\bibfnamefont{M.~G.} \bibnamefont{Boshier}},
  \bibinfo{author}{\bibfnamefont{H.~L.} \bibnamefont{Grossman}},
  \bibinfo{author}{\bibfnamefont{K.~P.} \bibnamefont{Zetie}}, \bibnamefont{and}
  \bibinfo{author}{\bibfnamefont{E.~A.} \bibnamefont{Hinds}},
  \bibinfo{journal}{Phys. Rev. Lett.} \textbf{\bibinfo{volume}{75}},
  \bibinfo{pages}{629} (\bibinfo{year}{1995}).

\bibitem[{\citenamefont{Sidorov et~al.}(1996)\citenamefont{Sidorov, McLean,
  Rowlands, Lau, Murphy, Walkiewicz, Opat, and Hannaford}}]{sidorov96}
\bibinfo{author}{\bibfnamefont{A.~I.} \bibnamefont{Sidorov}},
  \bibinfo{author}{\bibfnamefont{R.~J.} \bibnamefont{McLean}},
  \bibinfo{author}{\bibfnamefont{W.~J.} \bibnamefont{Rowlands}},
  \bibinfo{author}{\bibfnamefont{D.~C.} \bibnamefont{Lau}},
  \bibinfo{author}{\bibfnamefont{J.~E.} \bibnamefont{Murphy}},
  \bibinfo{author}{\bibfnamefont{M.}~\bibnamefont{Walkiewicz}},
  \bibinfo{author}{\bibfnamefont{G.~I.} \bibnamefont{Opat}}, \bibnamefont{and}
  \bibinfo{author}{\bibfnamefont{P.}~\bibnamefont{Hannaford}},
  \bibinfo{journal}{Quantum Semiclass. Opt.} \textbf{\bibinfo{volume}{8}},
  \bibinfo{pages}{713} (\bibinfo{year}{1996}).

\bibitem[{\citenamefont{Rosenbusch et~al.}(2000)\citenamefont{Rosenbusch, Hall,
  Hughes, Saba, and Hinds}}]{rosen2000}
\bibinfo{author}{\bibfnamefont{P.}~\bibnamefont{Rosenbusch}},
  \bibinfo{author}{\bibfnamefont{B.~V.} \bibnamefont{Hall}},
  \bibinfo{author}{\bibfnamefont{I.~G.} \bibnamefont{Hughes}},
  \bibinfo{author}{\bibfnamefont{C.~V.} \bibnamefont{Saba}}, \bibnamefont{and}
  \bibinfo{author}{\bibfnamefont{E.~A.} \bibnamefont{Hinds}},
  \bibinfo{journal}{Phys. Rev. A.} \textbf{\bibinfo{volume}{61}},
  \bibinfo{pages}{031404(R)} (\bibinfo{year}{2000}).

\bibitem[{\citenamefont{G{\"{u}}nther et~al.}(2005)\citenamefont{G{\"{u}}nther,
  Kraft, Kemmler, Koelle, Kleiner, Zimmermann, and Fort{\'{a}}gh}}]{gunther05}
\bibinfo{author}{\bibfnamefont{A.}~\bibnamefont{G{\"{u}}nther}},
  \bibinfo{author}{\bibfnamefont{S.}~\bibnamefont{Kraft}},
  \bibinfo{author}{\bibfnamefont{M.}~\bibnamefont{Kemmler}},
  \bibinfo{author}{\bibfnamefont{D.}~\bibnamefont{Koelle}},
  \bibinfo{author}{\bibfnamefont{R.}~\bibnamefont{Kleiner}},
  \bibinfo{author}{\bibfnamefont{C.}~\bibnamefont{Zimmermann}},
  \bibnamefont{and}
  \bibinfo{author}{\bibfnamefont{J.}~\bibnamefont{Fort{\'{a}}gh}},
  \bibinfo{journal}{Phys. Rev. Lett} \textbf{\bibinfo{volume}{95}},
  \bibinfo{pages}{170405} (\bibinfo{year}{2005}).

\bibitem[{\citenamefont{G{\"{u}}nther et~al.}(2007)\citenamefont{G{\"{u}}nther,
  Kraft, Zimmermann, and Fort{\'{a}}gh}}]{gunther07}
\bibinfo{author}{\bibfnamefont{A.}~\bibnamefont{G{\"{u}}nther}},
  \bibinfo{author}{\bibfnamefont{S.}~\bibnamefont{Kraft}},
  \bibinfo{author}{\bibfnamefont{C.}~\bibnamefont{Zimmermann}},
  \bibnamefont{and}
  \bibinfo{author}{\bibfnamefont{J.}~\bibnamefont{Fort{\'{a}}gh}},
  \bibinfo{journal}{Phys. Rev. Lett} \textbf{\bibinfo{volume}{98}},
  \bibinfo{pages}{140403} (\bibinfo{year}{2007}).

\bibitem[{\citenamefont{Gerritsma et~al.}(2007)\citenamefont{Gerritsma,
  Whitlock, Fernholz, Schlatter, Luigjes, Thiele, Goedkoop, and
  Spreeuw}}]{gerritsma:2007}
\bibinfo{author}{\bibfnamefont{R.}~\bibnamefont{Gerritsma}},
  \bibinfo{author}{\bibfnamefont{S.}~\bibnamefont{Whitlock}},
  \bibinfo{author}{\bibfnamefont{T.}~\bibnamefont{Fernholz}},
  \bibinfo{author}{\bibfnamefont{H.}~\bibnamefont{Schlatter}},
  \bibinfo{author}{\bibfnamefont{J.~A.} \bibnamefont{Luigjes}},
  \bibinfo{author}{\bibfnamefont{J.-U.} \bibnamefont{Thiele}},
  \bibinfo{author}{\bibfnamefont{J.~B.} \bibnamefont{Goedkoop}},
  \bibnamefont{and} \bibinfo{author}{\bibfnamefont{R.~J.~C.}
  \bibnamefont{Spreeuw}}, \bibinfo{journal}{Phys. Rev. A}
  \textbf{\bibinfo{volume}{76}}, \bibinfo{pages}{033408}
  (\bibinfo{year}{2007}).

\bibitem[{\citenamefont{Wang et~al.}(2005)\citenamefont{Wang, Whitlock,
  Scharnberg, Gough, Sidorov, McLean, and Hannaford}}]{wang2:05}
\bibinfo{author}{\bibfnamefont{J.~Y.} \bibnamefont{Wang}},
  \bibinfo{author}{\bibfnamefont{S.}~\bibnamefont{Whitlock}},
  \bibinfo{author}{\bibfnamefont{F.}~\bibnamefont{Scharnberg}},
  \bibinfo{author}{\bibfnamefont{D.~S.} \bibnamefont{Gough}},
  \bibinfo{author}{\bibfnamefont{A.~I.} \bibnamefont{Sidorov}},
  \bibinfo{author}{\bibfnamefont{R.~J.} \bibnamefont{McLean}},
  \bibnamefont{and}
  \bibinfo{author}{\bibfnamefont{P.}~\bibnamefont{Hannaford}},
  \bibinfo{journal}{J. Phys. D:Appl. Phys.} \textbf{\bibinfo{volume}{38}},
  \bibinfo{pages}{4015} (\bibinfo{year}{2005}).

\bibitem[{\citenamefont{Hall et~al.}(2006)\citenamefont{Hall, Whitlock,
  Scharnberg, Hannaford, and Sidorov}}]{whitlock:06}
\bibinfo{author}{\bibfnamefont{B.~V.} \bibnamefont{Hall}},
  \bibinfo{author}{\bibfnamefont{S.}~\bibnamefont{Whitlock}},
  \bibinfo{author}{\bibfnamefont{F.}~\bibnamefont{Scharnberg}},
  \bibinfo{author}{\bibfnamefont{P.}~\bibnamefont{Hannaford}},
  \bibnamefont{and} \bibinfo{author}{\bibfnamefont{A.~I.}
  \bibnamefont{Sidorov}}, \bibinfo{journal}{J. Phys. B: At. Mol. Opt. Phys.}
  \textbf{\bibinfo{volume}{37}}, \bibinfo{pages}{27} (\bibinfo{year}{2006}).

\bibitem[{\citenamefont{Perrin et~al.}(2006)\citenamefont{Perrin, Colombe,
  Mercier, Lorent, and Henkel}}]{perrin06}
\bibinfo{author}{\bibfnamefont{H.}~\bibnamefont{Perrin}},
  \bibinfo{author}{\bibfnamefont{Y.}~\bibnamefont{Colombe}},
  \bibinfo{author}{\bibfnamefont{B.}~\bibnamefont{Mercier}},
  \bibinfo{author}{\bibfnamefont{V.}~\bibnamefont{Lorent}}, \bibnamefont{and}
  \bibinfo{author}{\bibfnamefont{C.}~\bibnamefont{Henkel}},
  \bibinfo{journal}{J. Phys. B: At. Mol. Opt. Phys.}
  \textbf{\bibinfo{volume}{39}}, \bibinfo{pages}{4649} (\bibinfo{year}{2006}).

\end{thebibliography}

\end{document}